\documentclass[aps,prd,twocolumn,showpacs,amsmath,nofootinbib]{revtex4-1}

\usepackage{color}
\newcommand{\beqa}{\begin{eqnarray}}
\newcommand{\eeqa}{\end{eqnarray}}
\usepackage{float}
\usepackage{epsfig}
\usepackage{graphicx}
\usepackage{latexsym}
\usepackage{amsmath}
\usepackage{mathrsfs}
\usepackage{dcolumn}
\usepackage{bm}%

\begin{document}

\title{Impact of radiation from primordial black holes on the 21-cm angular-power spectrum in the dark ages}

\author{Yupeng Yang}

\affiliation{School of Physics and Physical Engineering, Qufu Normal University, Qufu, Shandong, 273165, China}

\begin{abstract}
We investigate the impact of radiation from primordial black holes (PBHs), in the mass range of 
$10^{15} \lesssim M_{\rm PBH} \lesssim 10^{17}\rm g$ and $10^{2} \lesssim M_{\rm PBH} \lesssim 10^{4}M_{\odot}$, on the 21-cm angular-power spectrum in the dark ages. PBHs in the former mass range affect the 21-cm angular-power spectrum through the evaporation known 
as Hawking radiation, while the radiation from the accretion process in the latter mass range. In the dark ages, radiation from PBHs can increase the ionization fraction 
and temperature of the intergalactic medium, change the global 21-cm differential brightness temperature and then affect the 21-cm 
angular-power spectrum. Taking into account the effects of PBHs, 
we find that in the dark ages, $30 \lesssim z \lesssim 100$, the amplitude of the 21-cm angular-power spectrum 
is decreased depending on the mass and mass fraction of PBHs. 
We also investigate the potential constraints on the mass fraction of PBHs 
in the form of dark matter for the future radio telescope in lunar orbit or on the far side surface of the Moon.

\end{abstract}

\maketitle

\section{introduction} 
In the standard cosmological model, dark matter makes up about  $27\%$ of the Universe~\cite{Planck:2018vyg}. 
Although many astronomical observations have confirmed the existence of dark matter, its nature has so far been unknown. 
Among the many dark matter models, weakly interacting massive particles (WIMPs) 
is the most important one~\cite{Bertone:2004pz,Jungman:1995df}. 
However, so far, all relevant experiments to detect WIMPs have not found any signs of them. 
Other dark matter models, such as primordial black holes (PBHs), have attracted extensive attention again
~\cite{Zou:2022wtp,carr,Bird:2022wvk,Carr:2021bzv,Wu:2021gtd,Flores:2021jas,Zhang:2021vak,Villanueva-Domingo:2021spv,Hasinger:2020ptw,Zhang:2021mth,Belotsky:2014kca,Laha:2019ssq,Dasgupta:2019cae,Laha:2020ivk,Laha:2020vhg}. 
Recently, the gravitational waves generated by the merger of black holes detected by LIGO/Virgo 
may be partly caused by PBHs~\cite{Bird:2016dcv,Clesse:2020ghq,Deng:2021ezy,Franciolini:2021tla,Chen:2021nxo,Hutsi:2020sol,Ashoorioon:2022raz}.

PBHs can be formed by the collapse of large density perturbation existing 
in the early Universe and their masses spread a wide range (see, e.g., Refs.~\cite{Carr:2005zd,carr,Khlopov:2008qy}). 
A PBH smaller than $M_{\rm PBH} \sim 10^{17}\rm g$ loses mass through evaporation 
due to Hawking radiation~\cite{Josan:2009,carr,Tashiro:2008sf,pbhs_emit_1,pbhs_emit_2,pbhs_emit_3,Kohri:2014lza,
Ray:2021mxu}. 
A massive PBH with mass $M_{\rm PBH} \gtrsim 10^{2}M_{\odot}$ radiates energy in the process of accretion
~\cite{Ricotti:2007jk,Ricotti:2007au,Poulin:2017bwe,Ali-Haimoud:2016mbv,Yang:2021agk,Carr:2020erq,DAgostino:2022ckg}. 
The extra energy injection from PBHs can affect the evolution of the intergalactic medium (IGM). 
The changes in the thermal history of the IGM will be reflected in, e.g., the anisotropy of the cosmic microwave background (CMB) 
and the global 21-cm signal
~\cite{Mukhopadhyay:2022jqc,Cang:2021owu,Yang:2021idt,Natwariya:2021xki,Mittal:2021egv,Tashiro:2021xnj,Yang:2020zcu,Chen:2016pud,mack_21,Cang:2020aoo,Mena:2019nhm,Saha:2021pqf}. 

Recently, the Experiment to Detect the Global Epoch of Reionization Signature reported 
the detection of the global 21-cm signal centered at redshift $z\sim17$ with an amplitude twice as large as expected~\cite{edges-nature}. 
Although this result needs to be further verified by other experiments, 
the observation can be used to study the related properties of PBHs, such as limiting their mass fraction. 
According to the theory, there are also 21-cm absorption signals in the dark ages of the Universe ($30\lesssim z\lesssim 100$)
~\cite{Pritchard:2011xb,Furlanetto:2006jb}, and these radio signals have been redshifted to 
the low frequency range ($14\lesssim\nu_{21}\lesssim 46$MHz). 
The Earth's ionosphere makes it impossible to detect these low-frequency signals from the Earth. 
Radio telescopes in orbit around the moon or on the far side of the moon have been proposed to avoid 
the influence of the ionosphere~\cite{2019arXiv190710853C,Burns:2021ndk,Burns_2020,Burns:2021pkx,2017arXiv170200286P,Chen:2019xvd,Shi:2022zdx}. 
In Ref.~\cite{Yang:2021agk}, the authors have investigated the effect of PBH accretion radiation on the global 21-cm signal 
in the dark ages, and explored the ability of future radio telescopes to limit the mass fraction of PBHs for 
the mass range of $10\lesssim M_{\rm PBH}\lesssim 10^{4}M_{\odot}$. Although the resulting constraints are 
not the strongest, they are still competitive with that of the lower redshift period because 
the dark ages are less affected by the formation of cosmic structures. 
Similar to the anisotropy of the cosmic microwave background, the 21-cm signals can also be studied using the angular-power spectrum~\cite{21ps,Loeb:2003ya,Santos:2004ju,Zaldarriaga:2003du}. 
The influence of dark matter annihilation on the 21-cm angular-power spectrum 
in the cosmic dawn has been studied in, e.g., Ref.~\cite{PhysRevD.80.043529}. 

In this paper, we focus on the influence of PBHs on the 21-cm angular-power spectrum in the dark ages. 
We mainly investigate the radiation from the evaporation and accretion process of PBHs, 
corresponding to the mass range of $10^{15}\lesssim M_{\rm PBH} \lesssim 10^{17}\rm g$ and 
$10^{2} \lesssim M_{\rm PBH}\lesssim 10^{4}M_{\odot}$, 
respectively~\footnote{A PBH with mass $M_{\rm PBH}\lesssim 10^{15}\rm g$ has a shorter lifetime than the age of the Universe~\cite{carr}. 
Here we only consider PBH with mass greater than $10^{15}\rm g$. It should be pointed out that 
the lower mass PBH can still affect the 21-cm angular-power spectrum in the dark ages.}. 
In view of a future extraterrestrial radio telescope, we investigate the ability of future detection of 
the 21-cm angular-power spectrum to limit the abundance of PBHs.

This paper is organized as follows. In Sec. II we investigate the influence of PBHs on the thermal history of the IGM 
and the global 21-cm signal in the dark ages. The 21-cm angular-power spectrum including PBHs and 
the future potential upper limits on the abundance of PBHs are discussed in Sec. III. 
The conclusions are given in Sec. IV. Throughout the paper we will use the cosmological parameters from Planck-2018 results
~\cite{Planck:2018vyg}.


\section{The global 21-cm signal in the dark ages including PBHs}
\subsection{The thermal history of the IGM including PBHs}
The changes in the thermal history of the Universe due to the injection of extra energy have been investigated 
by previous works (see, e.g., Refs.~\cite{xlc_decay,lz_decay,Slatyer:2015kla}). Here we review the main points and one can refer to, e.g., Refs.~\cite{xlc_decay,Slatyer:2015kla} for more 
details. 

The interactions between the particles emitted from PBHs with that existing in the Universe result 
in the changes of the thermal history of the IGM. Taking into account the effects of heating, ionization, and excitation, 
the changes of the degree of ionization ($x_e$) and the temperature of IGM ($T_k$) with the redshift 
are governed by the following equations~\cite{xlc_decay,lz_ann}:

\beqa
(1+z)\frac{dx_{e}}{dz}=\frac{1}{H(z)}\left[R_{s}(z)-I_{s}(z)-I_{\rm PBH}(z)\right],
\label{eq:xe}
\eeqa

\beqa
(1+z)\frac{dT_{k}}{dz}=&&\frac{8\sigma_{T}a_{R}T^{4}_{\rm CMB}}{3m_{e}cH(z)}\frac{x_{e}(T_{k}-T_{\rm CMB})}{1+f_{\rm He}+x_{e}}
\nonumber \\ 
&&-\frac{2}{3k_{B}H(z)}\frac{K_{\rm PBH}}{1+f_{\rm He}+x_{e}}+2T_{k}, 
\label{eq:tk}
\eeqa
where $R_{s}(z)$ and $I_{s}(z)$ are the recombination and ionization rate for the case with no PBHs, respectively. 
The ionization and heating rate caused by PBHs can be written as follows~\cite{mnras,yinzhema,DM_2015,Yang:2021idt}: 

\beqa
I_{\rm PBH} = f_{i}(z)\frac{1}{n_b}\frac{1}{E_{0}}\frac{{\rm d}E}{{\rm d}V{\rm d}t}\bigg|_{\rm PBH} 
\label{eq:I}
\eeqa
\beqa
K_{\rm PBH} = f_{h}(z)\frac{1}{n_b}\frac{{\rm d}E}{{\rm d}V{\rm d}t}\bigg|_{\rm PBH} 
\label{eq:K}
\eeqa
where $n_b$ is the number density of baryon. $E_0$ stands for the ground state energy of the hydrogen atom. 
$f(z)$ corresponds to the energy fraction injected into the IGM for ionization, heating and exciting, respectively. 
It has been studied in detail, e.g., Refs.~\cite{energy_function,Slatyer:2015kla,Poulin:2017bwe}, 
and we use the public code ExoCLASS~\cite{exoclass,class} to calculate $f(z)$ numerically.

For evaporating PBHs, the energy injection rate per unit volume is given by~\cite{yinzhema,prd-2020}

\beqa
\frac{{\rm d}E}{{\rm d}V{\rm d}t}\bigg|_{\rm PBH,eva} =f_{\rm pbh}\frac{\rho_{\rm DM}}{M_{\rm PBH}}\frac{{\rm d}M_{\rm PBH}}{{\rm d}t},
\eeqa 
where $f_{\rm pbh}=\rho_{\rm PBH}/\rho_{\rm DM}$. Here we have adopted a monochromatic PBH mass function 
for our calculations. The mass-loss rate of a black hole is~\cite{carr,Josan:2009}

\beqa
\frac{dM_{\rm BH}}{dt}=-5.34\times 10^{25}f(M_{\rm BH})\left(\frac{M_{\rm BH}}{\rm g}\right)^{-2}{\rm g}\rm s^{-1}
\eeqa
where $f(M_{\rm BH})$ is the number of particle species emitted directly and we have used the formula 
given in Ref.~\cite{Tashiro:2008sf}. 

For accreting PBHs, the energy injection rate per unit volume can be written as~\cite{Poulin:2017bwe,Ali-Haimoud:2016mbv} 

\beqa
\frac{{\rm d}E}{{\rm d}V{\rm d}t}\bigg|_{\rm PBH,acc} =f_{\rm pbh}\frac{\rho_{\rm DM}}{M_{\rm PBH}}L_{\rm acc, PBH},
\eeqa
where $L_{\rm acc, PBH}$ is the accretion luminosity, which is proportional to the 
Bondi-Hoyle rate $\dot M_{\rm HB}$~\cite{Poulin:2017bwe}:

\beqa
L_{\rm acc, PBH} = \epsilon \dot{M}_{\rm HB}c^2,
\eeqa
where $\epsilon$ is the radiative efficiency depending on the accretion details. 
The authors of~\cite{Ali-Haimoud:2016mbv} made a detailed analysis of the accretion process of PBHs, 
finding $\epsilon=10^{-5}~(10^{-3})\dot{m}$ for collisional ionization (photoionization). 
Here we use $\epsilon=10^{-5}\dot{m}$ for our calculations, corresponding to the conservative case. 
$\dot{m}$ is the dimensionless Bondi-Hoyle accretion rate, which is in the form of the Eddington luminosity $L_{\rm Edd}$ as 
$\dot{m} = \dot{M}_{\rm HB}c^{2}/L_{\rm Edd}$.


\begin{figure}[htb]
\centering
\includegraphics[width=0.48\linewidth]{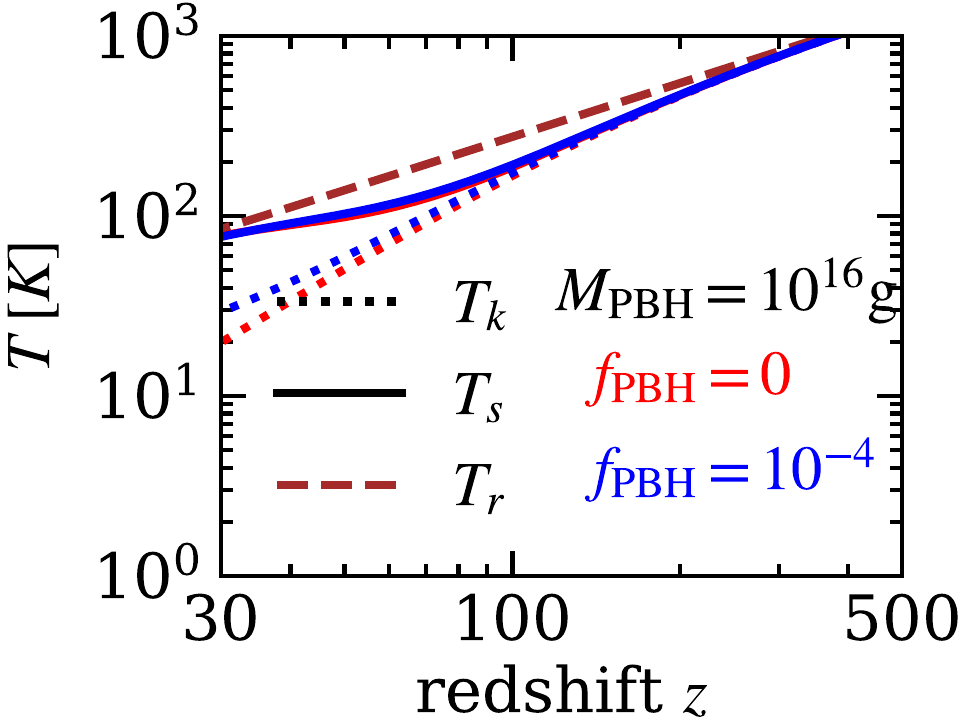}
\hfill
\includegraphics[width=0.48\linewidth]{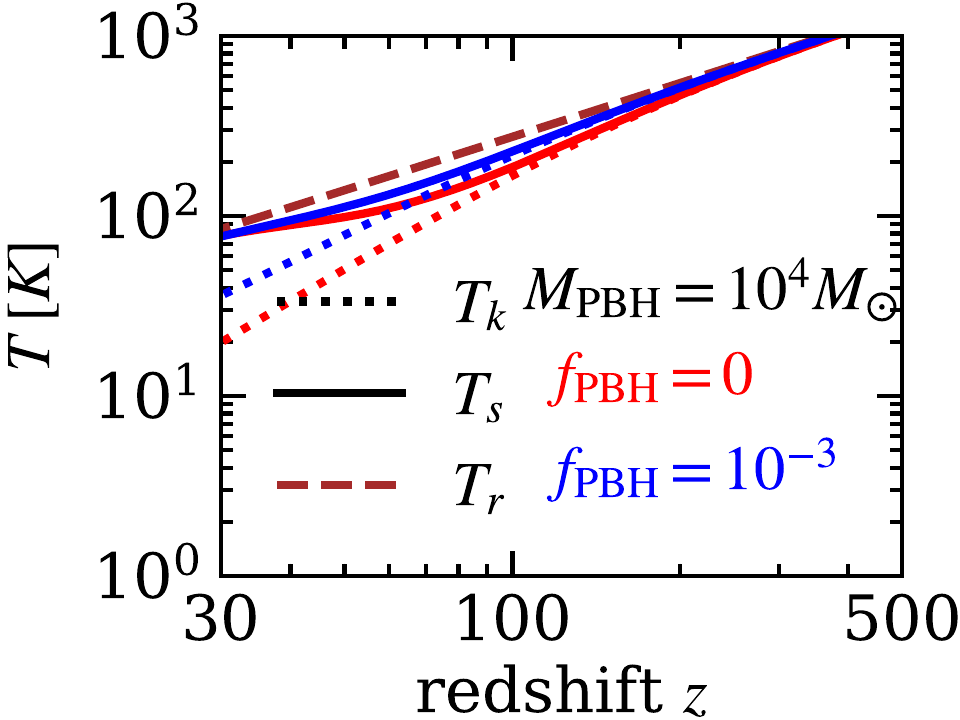}
\caption{The changes of the IGM temperature ($T_k$, dotted lines) and the spin temperature ($T_s$, solid lines) with redshift. 
Left: evaporating PBH with mass $M_{\rm PBH}=10^{16}\rm g$ and mass fraction $f_{\rm PBH}=10^{-4}$. 
Right: accreting PBH with mass $M_{\rm PBH}=10^{4}M_{\odot}$ and mass fraction $f_{\rm PBH}=10^{-3}$. 
For comparison, we also show the plots for the case with no PBH $f_{\rm PBH}=0$ (red lines). The CMB temperature ($T_r$) is also 
shown (brown dashed line).}
\label{fig:tem}
\end{figure}


In order to get the thermal history of the IGM for the case with PBHs, 
we have modified the public code RECFAST in CAMB\footnote{https://camb.info/} to solve the differential equations
~(\ref{eq:xe}) and (\ref{eq:tk}) numerically~\cite{mnras,DM_2015,xlc_decay,lz_decay,prd-2020,yinzhema}. The changes of $T_k$ with redshift are shown in Fig.~\ref{fig:tem}. 
In general, the injection of extra energy from PBHs raises the temperature of the IGM, 
and these effects are more pronounced at lower redshifts.


\subsection{The global 21-cm signal including PBHs}

Here we review the main issues about the global 21-cm signal. 
For more details and in-depth discussion, one can refer to, e.g., Refs.~\cite{Pritchard:2011xb,Furlanetto:2006jb} and references therein. 

The global 21-cm signal is usually described by the differential brightness temperature $\delta T_{21}$. 
Relative to the CMB background, $\delta T_{21}$ can be written as follows~\cite{Cumberbatch:2008rh,Ciardi:2003hg,prd-edges}: 

\beqa
\delta T_{21} =~&&26(1-x_e)\left(\frac{\Omega_{b}h}{0.02}\right)\left[\frac{1+z}{10}\frac{0.3}{\Omega_{m}}\right]^{\frac{1}{2}} \nonumber \\
&&\times \left(1-\frac{T_{\rm CMB}}{T_s}\right)\rm mK,
\label{eq:t_21}
\eeqa
where $\Omega_{b}$ and $\Omega_{m}$ are the density parameters of baryonic matter and dark matter, respectively. 
$h$ is the reduced Hubble constant. $T_s$ is the spin temperature defined as~\cite{Pritchard:2011xb,Furlanetto:2006jb} 

\beqa
\frac{n_1}{n_0}=3~\mathrm{exp}\left(-\frac{0.068K}{T_s}\right),
\eeqa
where $n_0$ and $n_1$ are the number densities of hydrogen atoms in triplet and singlet states, respectively. 
Specifically, the spin temperature can be written in the form of a weighted mean of the CMB temperature 
($T_{\rm CMB}$) and the IGM temperature ($T_k$)~\cite{Cumberbatch:2008rh,binyue}

\beqa
T_{s} = \frac{T_{\rm CMB}+(y_{\alpha}+y_{c})T_{k}}{1+y_{\alpha}+y_{c}},
\eeqa
where $y_{\alpha}$ corresponds to the Wouthuysen-Field effect and we use the formula given in, 
e.g., Refs.~\cite{binyue,mnras,Kuhlen:2005cm}: 

\beqa
y_{\alpha} = \frac{P_{10}}{A_{10}}\frac{0.068}{T_{k}}{\rm e}^{{-\frac{0.3\sqrt{1+z}}{T_{k}^{2/3}}}}\left(1+\frac{0.4}{T_{k}}\right)^{-1},
\eeqa
where $A_{10}=2.85\times 10^{-15}s^{-1}$ is the Einstein coefficient of hyperfine spontaneous transition. 
$P_{10}$ is the radiative deexcitation rate due to Ly$\alpha$ photons
~\cite{Pritchard:2011xb,Furlanetto:2006jb}. Taking into account the collisions between hydrogen atoms and other particles, $y_c$ can be written as~\cite{binyue,prd-edges,Kuhlen:2005cm,Liszt:2001kh,epjplus-2}

\beqa
y_{c} = \frac{0.068(C_{\rm HH}+C_{\rm eH}+C_{\rm pH})}{A_{10}T_{k}},
\eeqa  
where $C_{\rm HH, eH, pH}$ are the deexcitation rates of collisions~\cite{prd-edges,epjplus-2,Kuhlen:2005cm,Liszt:2001kh}.


\begin{figure}[htb]
\centering
\includegraphics[width=0.48\linewidth]{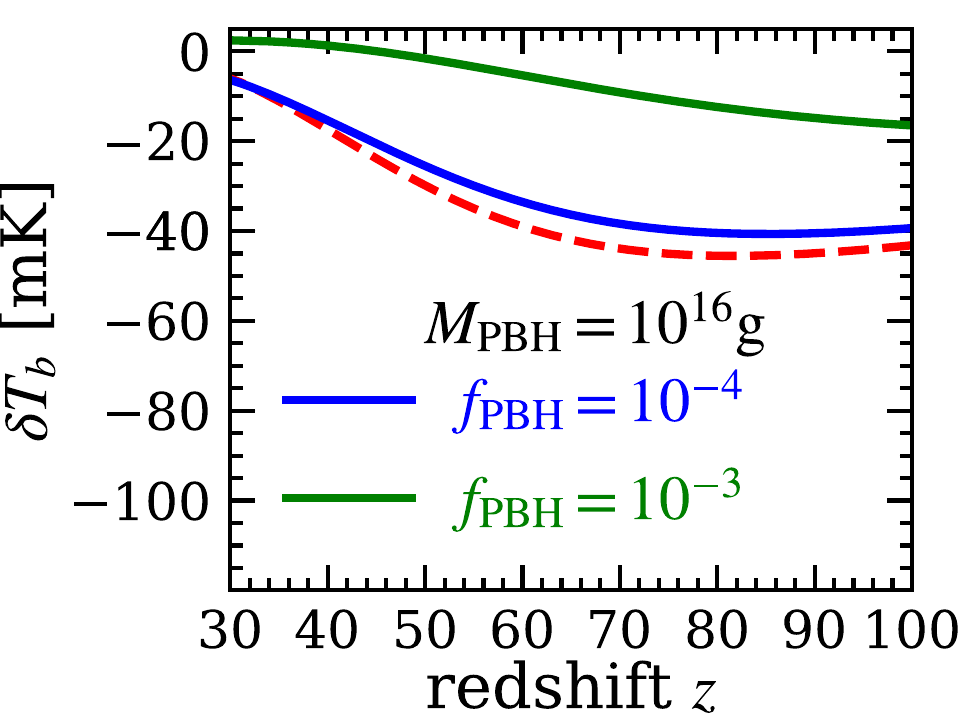}
\hfill
\includegraphics[width=0.48\linewidth]{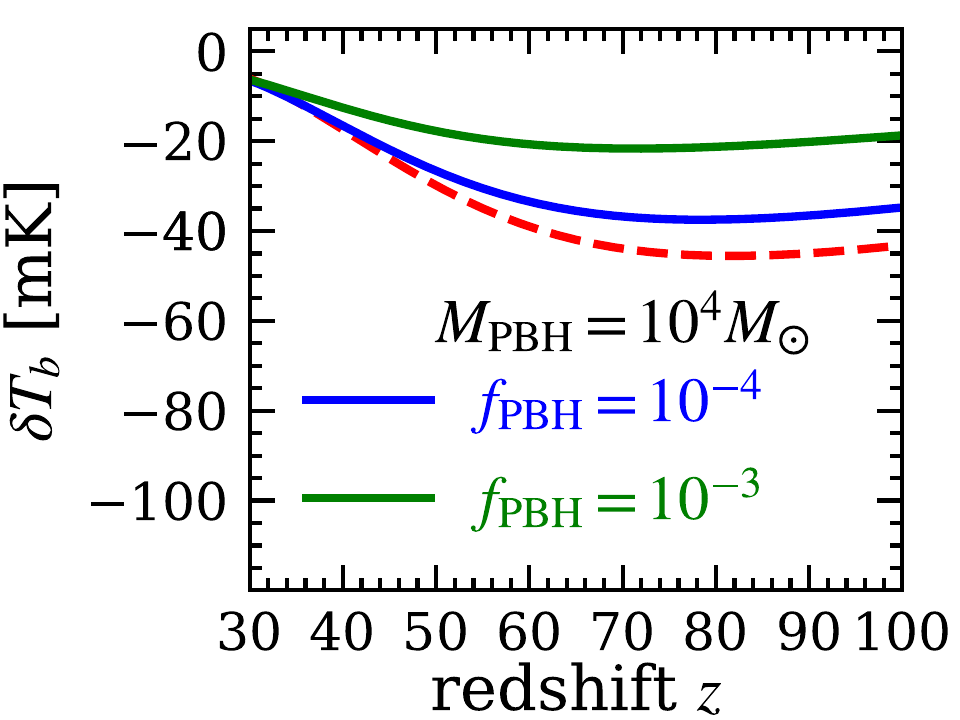}
\caption{The changes of the differential brightness temperature $\delta T_{21}$ with redshift for mass fraction 
$f_{\rm PBH}=10^{-4}$ (blue solid line) and $f_{\rm PBH}=10^{-3}$ (green solid line). 
Left: evaporating PBH with mass $M_{\rm PBH}=10^{16}\rm g$. 
Right: accreting PBH with mass $M_{\rm PBH}=10^{4}M_{\odot}$. The case with no PBH $f_{\rm PBH}=0$ 
is also shown (red dashed line).}
\label{fig:delta}
\end{figure}


The changes of the spin temperature $T_s$ with redshift are shown in Fig.~\ref{fig:tem}. 
It can be seen that $T_s$ becomes larger than that with no PBH, depending on the 
mass and mass fraction of PBH. The changes of the differential brightness temperature $\delta T_{21}$ with redshift 
are shown in Fig.~\ref{fig:delta}. The amplitude of the 21-cm absorption signal is decreased due to the influence of PBH. 
For a larger mass fraction of PBH with a fixed mass, the emission signal appears as shown 
in Fig.~\ref{fig:delta} for $f_{\rm PBH}=10^{-3}$ with $M_{\rm PBH} = 10^{16}\rm g$.

\section{The 21-cm angular-power spectrum and upper limits on the mass fraction of PBHs}
\begin{figure}[htb]
\centering
\includegraphics[width=0.48\linewidth]{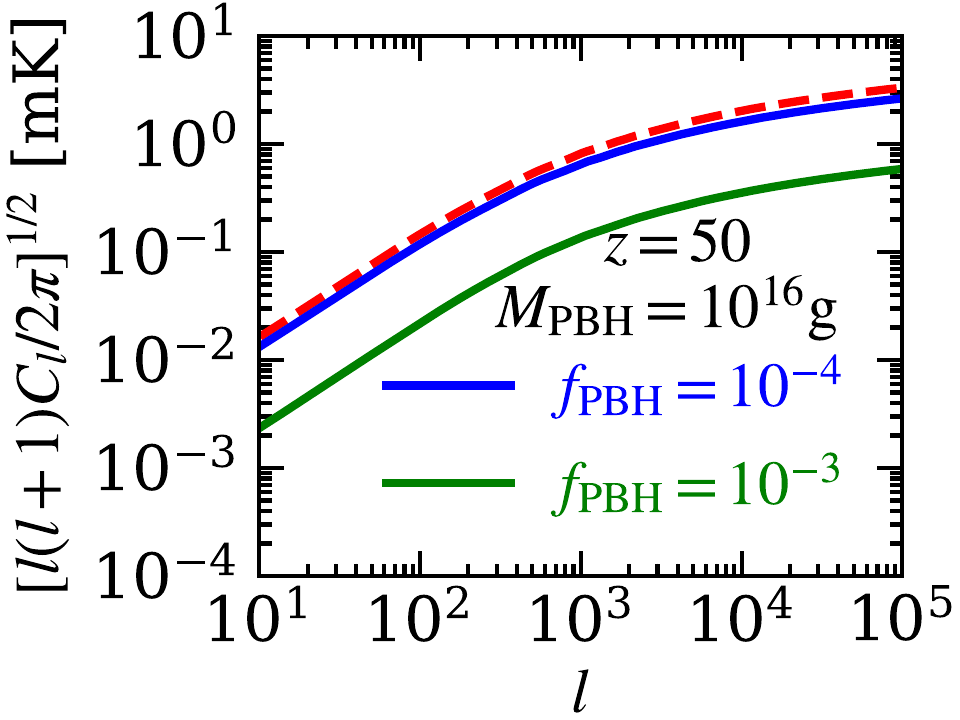}
\hfill
\includegraphics[width=0.48\linewidth]{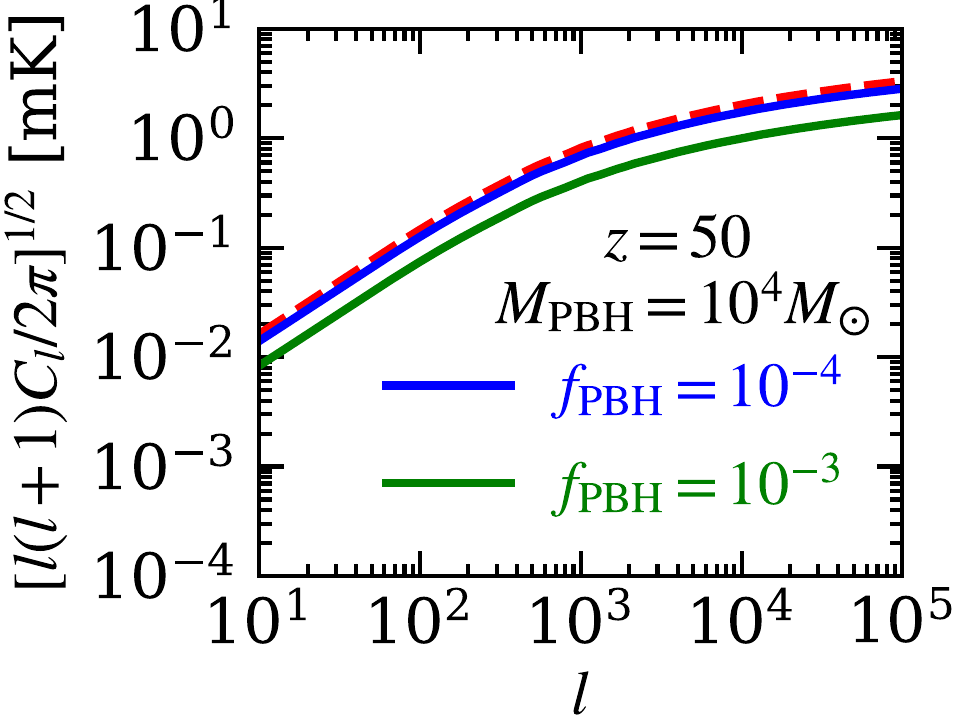}
\caption{The 21-cm angular-power spectrum at redshift $z=50$ including PBH for mass fraction 
$f_{\rm PBH}=10^{-4}$ (blue solid line) and $f_{\rm PBH}=10^{-3}$ (green solid line). 
Left: evaporating PBH with mass $M_{\rm PBH}=10^{16}\rm g$. 
Right: accreting PBH with mass $M_{\rm PBH}=10^{4}M_{\odot}$.}
\label{fig:ps}
\end{figure}

Similar to the CMB anisotropy, the fluctuations of $\delta T_{21}$ can also be described by the 21-cm 
angular-power spectrum, which can be calculated by using a standard Boltzmann
code. The calculation details of 21-cm angular-power spectrum 
can be found in Ref.~\cite{21ps} and the numerical code is available in CAMB. Here we have used the public code CAMB for our calculations, which has been used in the previous 
section to investigate the thermal history of the IGM including the effects of PBHs. The 21-cm angular-power spectrum 
at redshift $z=50$ is shown in Fig.~\ref{fig:ps}. 


\begin{figure}[htb]
\centering
\includegraphics[width=0.48\linewidth]{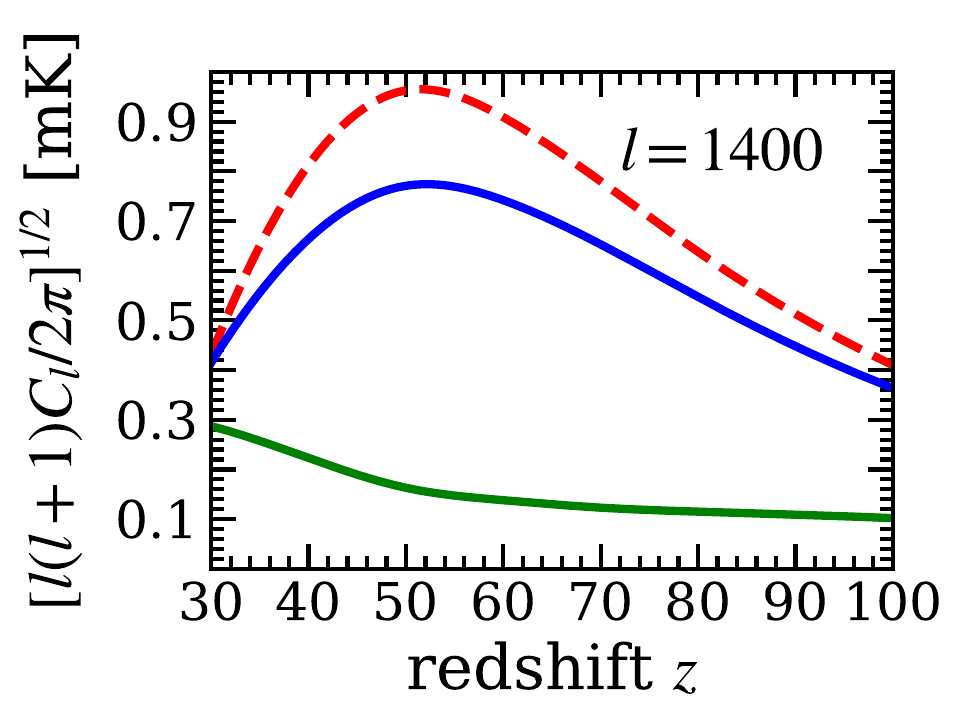}
\hfill
\includegraphics[width=0.48\linewidth]{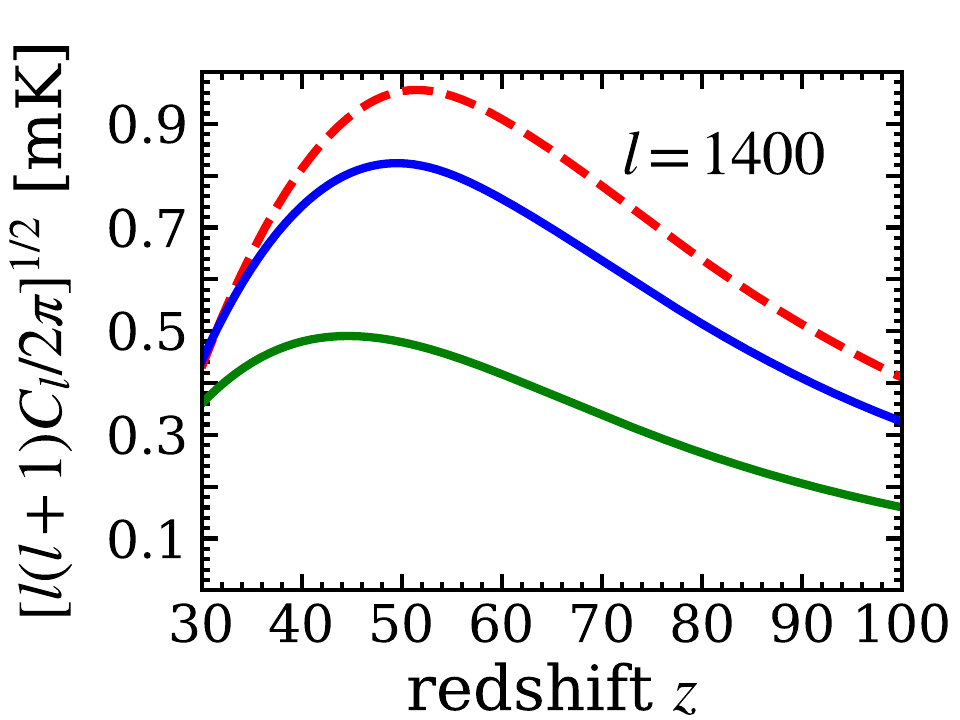}
\caption{The evolution of the 21-cm angular-power spectrum with redshift in the dark ages ($30\lesssim z\lesssim 100$) 
for a scale $l=1400$. The line style is the same as in Figs.~\ref{fig:delta} and~\ref{fig:ps}. 
Left: evaporating PBH with mass $M_{\rm PBH}=10^{16}\rm g$. 
Right: accreting PBH with mass $M_{\rm PBH}=10^{4}M_{\odot}$.}
\label{fig:zps}
\end{figure}

For the case with no PBH, the amplitude of the 21-cm  angular-power spectrum is about 
1 $\sim$ 3mK for $l\sim 10^{3}-10^{5}$. Since the angular-power spectrum is roughly proportional to 
$\left |\delta T_{21}\right|$~\cite{21ps,Furlanetto:2006jb}, therefore, for the case with PBH, the angular-power spectrum is decreased depending on 
the mass and mass fraction of PBH. In Fig.~\ref{fig:zps}, we also show the 21-cm angular-power spectrum for a 
scale $l=1400$ in the redshift range $30\lesssim z\lesssim 100$. For the case with no PBH, the largest amplitude 
of the angular-power spectrum appears at redshift $z\sim 50$~\footnote{Similar results can also be found for other scales~\cite{Pritchard:2011xb,Furlanetto:2006wp}.}. Including the effects of PBH, the largest amplitude 
shifts to the lower redshifts.

The 21-cm signal ($\nu_{21} =1421 $ MHz) from the redshift range $30\lesssim z\lesssim 100$ has been redshifted into 
the frequency range $14\lesssim \nu_{21} \lesssim 46$MHz. Due to the influence of the Earth's ionosphere, 
it is difficult to detect these low frequency signals from the Earth. 
A radio telescope, either in lunar orbit or on the far side surface of the Moon, has been proposed to detect 
these radio signals~\cite{2019arXiv190710853C,Burns:2021ndk,Burns_2020,Burns:2021pkx,2017arXiv170200286P,Chen:2019xvd}. 
For a radio telescope, the uncertainty of the $C_{l}$ at a mutlipole $l$ is~\cite{Bernal:2017nec,Zaldarriaga:2003du,Kesden:2002ku}

\beqa
\sigma_{C_{l}}=\sqrt{\frac{2(C_{l}+C_{l}^{N})^{2}}{f_{\rm sky}(2l+1)}},
\eeqa  
where $C_{l}^{N}$ is the noise power spectrum~\cite{Bernal:2017nec,Zaldarriaga:2003du} 

\beqa
l^{2}C_{l}^{N}=\frac{(2\pi)^{2}T^{2}_{\rm sky}}{\Delta \nu t_{\rm obs}f^{2}_{\rm cover}}\left(\frac{l}
{l_{\rm max}}\right)^{2},
\eeqa
where $t_{\rm obs}$ is the observation time, $l_{\rm max}$ is the maximum multipole observable, 
and $f_{\rm cover}$ is the array covering factor. $T_{\rm sky}$ is the sky temperature.
For the low frequency range, $\nu < 100\rm MHz$, $T_{\rm sky}$ is dominated by the galactic 
synchrotron radiation background and scales as $\nu^{-\alpha}$
~\cite{Furlanetto:2006jb,Oberoi:2003zn,Jester_2009,Jester:2009dw,Platania:1997zn,
1982A&AS,1987MNRAS.225..307L}. A more detailed analysis of $T_{\rm sky}$ can be also found in, e.g., Ref.~\cite{planck_syn_index}. Here we adopted the approximated form given in Ref.~\cite{Jester:2009dw}:
\beqa
T_{\rm sky} = 16.3\times10^{6}\left(\frac{\nu}{2\rm MHz}\right)^{-2.53}\rm K.
\eeqa 
Another approximated form usually used is $T_{\rm sky} = 180\left(\nu/180\rm MHz\right)^{-2.6}\rm K$, 
see, e.g., Refs.~\cite{Furlanetto:2006jb,Valdes:2007cu}. 
Note that the differences between two formulas have a negligible impact on the estimation of our final results.

For a future radio telescope, e.g., on the lunar surface~\cite{Bernal:2017nec,farside}, 
with an array size $D \sim 300$km, the maximum multipole 
could reach $l_{\rm max}\sim 10^{5}$ at redshift $z\sim 50$. Therefore, for the array covering factor 
$f_{\rm cover}\sim 0.75$ and bandwidth $\Delta \nu \sim 50$MHz, 
the uncertainty of the $C_l$ at $z=50$ for $l=1400$ could be $\sigma_{C_{l}}\sim 0.02$mK 
for 1000 hours observation time and $f_{\rm sky} \sim 1$. Therefore, a large deviation of the 21-cm angular-power spectrum 
from the default case could be detected for the future radio telescope. On the other hand, 
future observations of the 21-cm angular-power spectrum can be used to put limits on the abundance of PBHs. 
Here we will make a simple study of the abundance of PBHs for the future detection, 
and more detailed studies are left for future work. 

\begin{figure}[htb]
\centering
\includegraphics[width=0.48\linewidth]{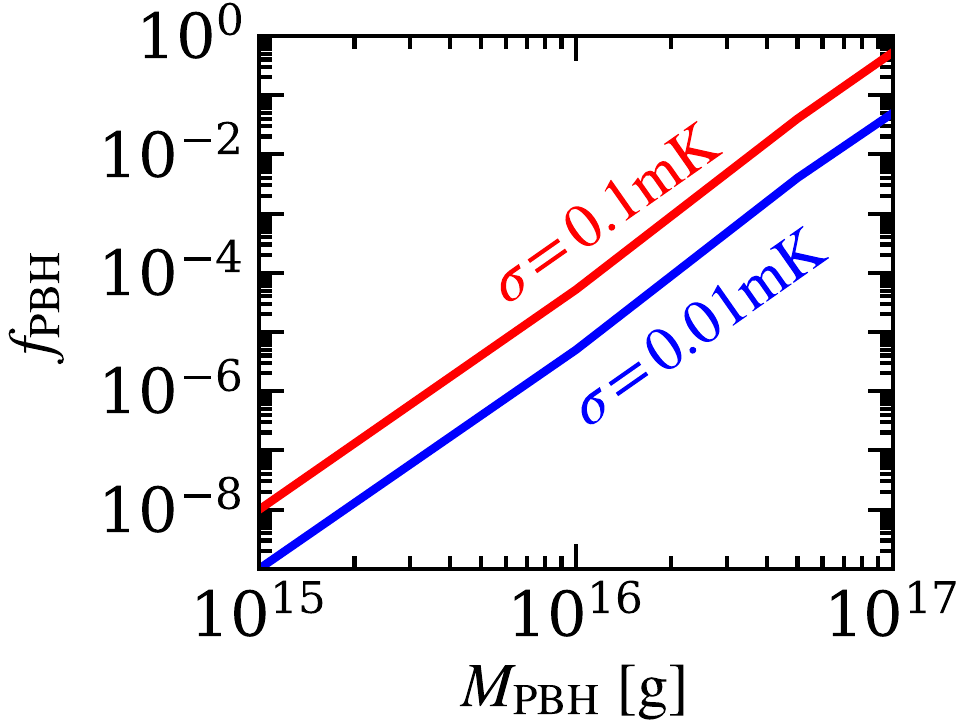}
\hfill
\includegraphics[width=0.48\linewidth]{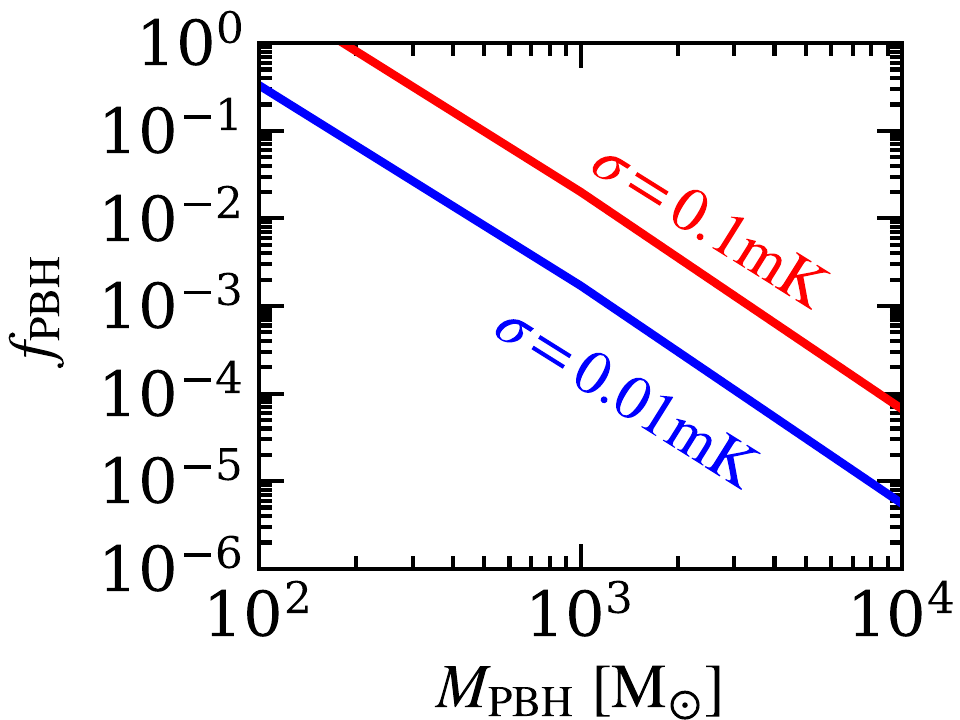}
\caption{Upper limits on the mass fraction of PBHs for future radio detection. 
Instead of focusing on the sensitivity of a specific radio telescope, we have set $\sigma_{C_{l}}$ = 0.1 (red line) 
and 0.01 mK (blue line) for our calculations. 
Left: evaporating PBH in the mass range of $10^{15}\lesssim M_{\rm PBH}\lesssim 10^{16}\rm g$. 
Right: accreting PBH in the mass range of $10^{2}\lesssim M_{\rm PBH}\lesssim 10^{4}M_{\odot}$.}
\label{fig:cons}
\end{figure}

Instead of focusing on the sensitivity of a specific radio telescope, we have set $\sigma_{C_{l}}=0.1$ and $0.01$mK for our calculations, 
which could be achieved in the future. Moreover, for simply, we have focused on the maximum sensitivity at 
a specific scale $l$ instead of all scales~\cite{Zaldarriaga:2003du}. 
By requiring the deviation of the 21-cm angular-power spectrum less than $\sigma_{C_{l}}$ for 
$l=1400$ at redshift $z=50$, we find the upper limits on the mass fraction of PBHs $f_{\rm PBH}$, which are shown in Fig.~\ref{fig:cons}. For evaporating PBHs, the strongest limit is $f_{\rm PBH}\sim 10^{-8}~(10^{-9})$ for $\sigma_{C_{l}}=0.1$~(0.01)$\rm mK$ for $M_{\rm PBH}\sim 10^{15}\rm g$. For accreting PBHs, the strongest limit is $f_{\rm PBH}\sim 6\times 10^{-5}~(10^{-6})$ 
for $\sigma_{C_{l}}=0.1$~(0.01)$\rm mK$ for $M_{\rm PBH}\sim 10^{4}M_{\odot}$. Note that these constraints are comparable to the existing ones~\cite{carr}. Since these limits are from the dark ages, where the influence of astrophysical factors 
is smaller than that in the later period, therefore, future extraterrestrial detection of the radio signal 
can give very competitive results for limiting the mass fraction of PBHs.


\begin{figure}[htb]
\centering
\includegraphics[width=0.48\linewidth]{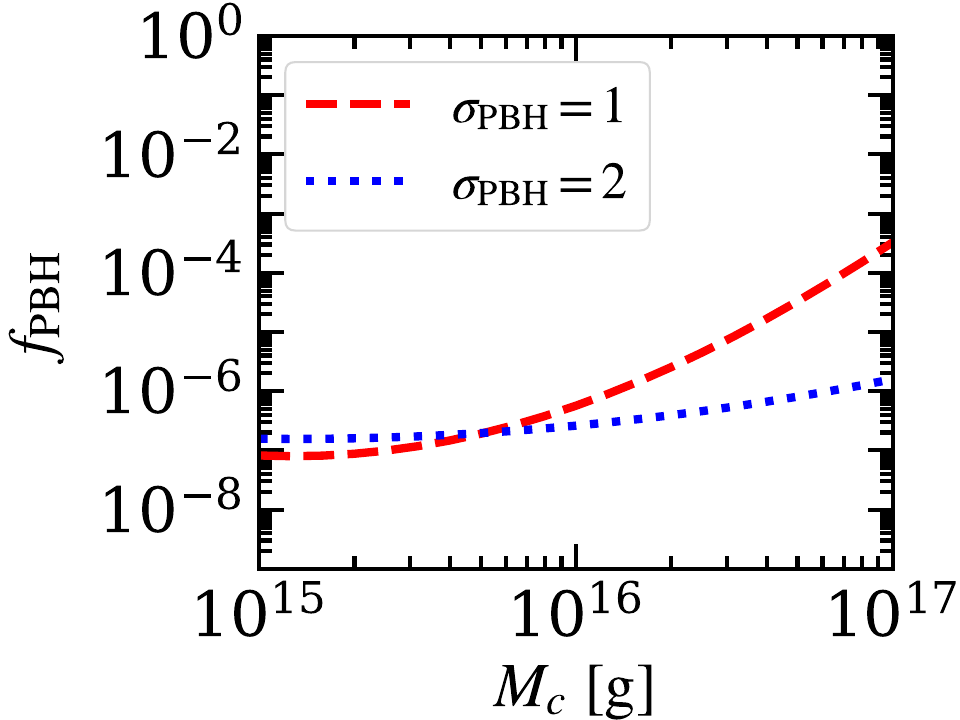}
\hfill
\includegraphics[width=0.48\linewidth]{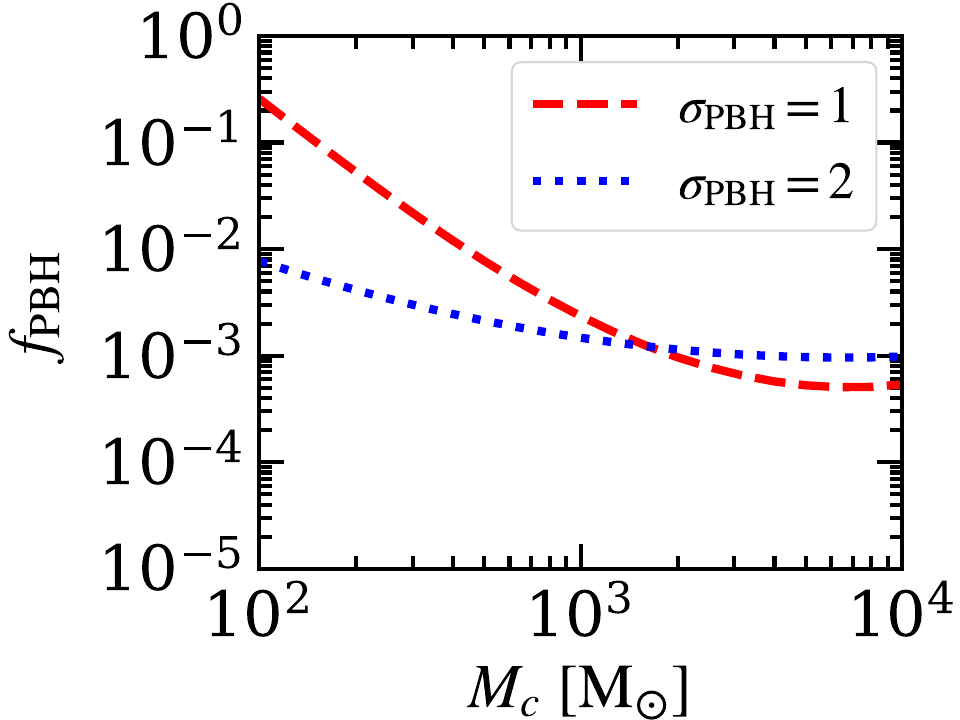}
\caption{Upper limits on the mass fraction of PBHs for extended mass function for 
$\sigma_{\rm PBH}=1$ (dashed line) and 2 (dotted line). Here we have set $\sigma_{C_{l}}$ = 0.1mK. 
Left: evaporating PBH in the mass range of $10^{15}\lesssim M_{\rm PBH}\lesssim 10^{16}\rm g$. 
Right: accreting PBH in the mass range of $10^{2}\lesssim M_{\rm PBH}\lesssim 10^{4}M_{\odot}$.
}
\label{fig:cons_exten}
\end{figure}

Although the monochromatic PBH mass function has been used usually, the extended 
mass function should be more realistic as predicted by many formation scenarios; see, e.g., Refs.
~\cite{mf_1,mf_2,mf_3,mf_4,mf_5,mf_6,mf_7,mf_8,mf_9}. Here we also investigate the constraints on $f_{\rm PBH}$ 
for extended PBH mass function. We consider one of the typical extended 
PBH mass functions, log-normal distribution~\cite{mf_7,Carr:2020xqk}, as follows:
\beqa
\Psi(M)=\frac{1}{\sqrt{2\pi}\sigma_{\rm PBH} M}\rm{exp}\left[-\frac{(logM-logM_{c})^{2}}{2\sigma^2_{\rm PBH}}\right].
\eeqa 
Using the constraints from the monochromatic PBH mass function, one can derive the upper limits on $f_{\rm PBH}$ 
for extended PBH mass distribution~\cite{mf_7,Cang:2020aoo,Cang:2021owu,Kuhnel:2017pwq},  
\beqa
f_{\rm PBH}\leq \left[\int dM\frac{\Psi(M,M_{c},\sigma_{\rm PBH})}{f_{\rm PBH}^{\prime}(M)}\right]^{-1},
\eeqa
where $f_{\rm PBH}^{\prime}(M)$ is the constraint for monochromatic PBH mass function shown in Fig.~\ref{fig:cons}. 
The upper limits on $f_{\rm PBH}$ for extended PBH mass function are shown in Fig.~\ref{fig:cons_exten}.
\footnote{Note that the extended mass function should be also included in the calculations 
of the energy injection into IGM~\cite{Cang:2020aoo,mf_9}. 
The final constraints should be different by several factors~\cite{Cang:2020aoo,yinzhema}.}

Another important issue for the influence of PBHs on the 21-cm angular-power 
spectrum is the Poissonian fluctuation in the number density of PBHs, 
which has an extra contribution to the standard matter power spectrum
~\cite{Villanueva-Domingo:2021cgh,Bernal:2017nec,Tashiro:2012qe,Cole:2019zhu,Afshordi:2003zb}. 
Compared with the effects of PBHs considered here, 
the amplitude of the 21-cm angular-power spectrum is increased due to Poissonian fluctuation. 
As shown in, e.g., 
Refs.~\cite{Bernal:2017nec,Cole:2019zhu}, Poissonian fluctuation 
has a large effect at lower redshifts ($z\lesssim 50 $) or on smaller scales ($l\gtrsim 10^{3}$), 
depending on the related parameters. At higher redshifts, the Poissonian contribution 
becomes smaller even on smaller scales~\cite{Bernal:2017nec,Cole:2019zhu}. 
Given the opposite effects of the radiation and Poissonian fluctuation of PBHs on the 
21-cm angular-power spectrum, it is expected that 
the future detection of power spectrum at higher redshifts can significantly reduce the 
impact of Poissonian contribution.

Many other factors can also affect the 21-cm angular-power spectrum~\cite{21ps}. 
Correspondingly, the detection of the 21-cm angular-power spectrum can be used to investigate those factors, 
such as the primordial power spectrum ($\mathcal{P_{R}}(k)$), which is related to the formation scenario of PBHs. 
The primordial power spectrum at large scales (small $k$) has been constrained by many observations, 
which is basically consistent with the scale invariant spectrum predicted by the popular inflation model
~\cite{cmb_2,lyman,large}. 
In general, the primordial power spectrum at small scales (large $k$) should be enhanced in order to form PBHs; 
see, e.g., 
Refs.~\cite{Cole:2019zhu,Germani:2017bcs,Byrnes:2018txb,Passaglia:2018ixg,Zhai:2022mpi,Raveendran:2022dtb,Cai:2022erk,
Heydari:2021qsr,Yi:2020cut,Unal:2020mts,Cai:2019bmk,Mishra:2019pzq,Kalaja:2019uju,Carrilho:2019oqg,Gao:2018pvq,
Zhou:2020kkf,Cai:2018tuh,Ashoorioon:2019xqc,Ashoorioon:2020hln}. 
Many different formation scenarios of PBHs, corresponding to the different forms of $\mathcal{P_{R}}(k)$ 
and matching the measurements on small scales, have been proposed. As shown in, e.g., Ref~\cite{Cole:2019zhu}, 
the 21-cm power spectrum (three dimensions) can be used to investigate these different scenarios 
since the 21-cm signal covers smaller scales and many more modes than those such as CMB 
measurements. In theory, the 21-cm angular-power spectrum can also be used to study different formation 
scenarios of PBHs. Compared with the scale invariant spectrum, 
the enhancement of $\mathcal{P_{R}}(k)$ at small scale (larger $k$) 
can result in the increasing of the 21-cm angular-power spectrum 
at small scale (large $l$)~\cite{Cole:2019zhu}. 
It is expected that these deviations from the standard power spectrum, corresponding to the 
different formation scenarios of PBHs, would also be examined by future detection 
of the 21-cm angular-power spectrum. We will conduct a detailed analysis of the relevant issues in future work.

Note that in obtaining the above limits, we have fixed other cosmological parameters except $f_{\rm PBH}$. 
On the other hand, the parameter $f_{\rm PBH}$ can affect the estimation of cosmological parameters~\cite{Loeb:2003ya}, 
and the details of these effects are complicated to analyze. 
Roughly, for example, the 21-cm angular-power spectrum is proportional to 
$\left |\delta T_{21}\right|$~\cite{21ps,Furlanetto:2006jb}, which is related to $\Omega_b$ as shown in Eq.~(\ref{eq:t_21}). Decreasing (increasing) the baryon density will result in lowering (boosting) 
the amplitude of 21-cm angular-power spectrum. Therefore, it is expected that $\Omega_b$ and $f_{\rm PBH}$ 
should be positively correlated in view of the 21-cm angular-power spectrum in the dark ages. Note that 
the analysis here is simple. 
A complete multiparameter analysis should be carried out by the statistical method such as Markov chain Monte Carlo, and we will perform these analyses in future work.

\section{CONCLUSIONS}

We have investigated the impact of PBHs on the thermal history of IGM and the 21-cm angular-power spectrum 
in the dark ages. Previous works mainly focused on the effects of the radiation 
from accreting PBHs on the global 21-cm signal and 21-cm power spectrum in 
the cosmic dawn and epoch of reionization. 
Here we have also focused on the evaporating PBHs in the mass range of 
$10^{15}\lesssim M_{\rm PBH}\lesssim 10^{16}\rm g$, besides the accreting PBHs in the mass range of 
$10^{2}\lesssim M_{\rm PBH}\lesssim 10^{4}M_{\odot}$. 
The radiation from PBHs results in increasing 
the gas and spin temperature compared with the case with no PBHs. The amplitude of the 21-cm absorption signal 
is decreased, and the emission signal appears for a larger mass fraction of PBHs. The fluctuations of the 21-cm differential 
brightness temperature can be described by the 21-cm angular-power spectrum. Taking into account the effects of PBHs, 
the 21-cm angular-power spectrum is decreased in the dark ages depending on the mass and mass fraction of PBHs. 
The peak value of the 21-cm angular-power spectrum appears at redshift $z\sim 50$ for the case with no PBHs, 
and shifts to the lower redshifts including PBHs. 

The 21-cm signals from the dark ages have been redshifted into the lower frequency $14\lesssim \nu_{21}\lesssim 46\rm MHz$ 
($30\lesssim z\lesssim 100$). 
It is difficult to detect these radio signals from the Earth due to the influence of Earth's ionosphere. 
Extraterrestrial radio telescopes, such as in lunar orbit or on the lunar surface, have been proposed. 
In view of the future radio telescopes, we have estimated the upper limits on the mass fraction of PBHs. 
Instead of focusing on a specific radio telescope, we have set the uncertainty $\sigma_{C_{l}}$ = 0.1 and 0.01mK, 
which can be achieved in the future, for our calculations. For a evaporating PBH with mass $M_{\rm PBH}\sim 10^{15}\rm g$, 
the upper limit is $f_{\rm PBH}\sim 10^{-8}~(10^{-9})$ for $\sigma_{C_{l}}=0.1~(0.01)\rm mK$. For a accreting PBH with mass 
$M_{\rm PBH}\sim 10^{4}~M_{\odot}$, the upper limit is $f_{\rm PBH}\sim 6\times 10^{-5}~(10^{-6})$ 
for $\sigma_{C_{l}}=0.1~(0.01)\rm mK$. Compared with the cosmic dawn and epoch of reionization, 
the dark ages is less affected by the astrophyscial factors. 
Therefore, the detection of the 21-cm signal or angular-power spectrum in the dark ages 
will be of great significance to reveal the relevant properties of PBHs.

\section{Acknowledgements}
Y. Yang thanks Dr. Bin Yue, Yan Gong and Yidong Xu for the helpful discussions and suggestions. 
This work is supported by the Shandong Provincial Natural Science Foundation (Grant No.ZR2021MA021). 
Y. Yang is supported in part by the Youth Innovations and Talents Project of Shandong Provincial Colleges and Universities (Grant No. 201909118).
\

\bibliographystyle{apsrev4-1}
\bibliography{refs}

\end{document}